\documentstyle[prl,aps,multicol,epsf]{revtex}
\tighten
\begin{document}
\draft

\title{Rectification of Fluctuations in an Underdamped Ratchet}

\author{Ya. M. Blanter and M. B\"uttiker}
\address{D\'epartement de Physique Th\'eorique, Universit\'e 
de Gen\`eve, CH-1211, Gen\`eve 4, Switzerland} 
\date{\today}
\maketitle

\begin{abstract}
We investigate analytically the motion of underdamped particles
subject to  a deterministic periodic potential and a periodic
temperature. Despite the fact that an underamped particle
experiences the temperature oscillation many times in its escape out
of a well and in its motion along the potential, a net directed
current linear in the friction constant is found. If both the
potential and the temperature modulation are sinusoidal with a phase
lag $\delta$, this current is proportional to $\sin \delta$.  
 
\end{abstract}


\begin{multicols}{2}

Recently, research on molecular motors\cite{JAP,motor}, which
apparently operate with an efficiency close to $kT$, has stimulated
interest in ratchets. Ratchets are commonly defined as systems which,
in the absence of a net force or macroscopic gradient, are able to 
produce a directed current through the rectification of noise\cite{Feynm}. 
Fundamentally ratchets are of interest as simple non-equilibrium systems. 
Some of the proposed models of ratchets consider particles in a periodic
potential and subject to a non-uniform periodic temperature profile
\cite{Buttiker,Kampen,Landauer1,Millonas}, or subject to a time-dependent
external force with zero average \cite{force}, which, in
particular, can be stochastic. Other models invoke
Brownian motion in fluctuating potentials \cite{fluct} or models of
Brownian particles which can be in several states with external
pumping of particles to one of the states \cite{Prost}. These works
treat overdamped  particles. Models in which inertia plays a role
either within classical or quantum dynamics have been considered only
recently \cite{inertia,quant1}, and the results available by now are
mostly numerical.

We present analytical results for the motion of {\em underdamped}
particles in a periodic potential and subject to a periodic
temperature modulation (see Fig.~1). In the particular case shown in
Fig.~1a, both the temperature modulation and the potential are
sinusoidal with the same microscopic period and a relative phase lag
$\delta$. We show that a directed current results which is
proportional to $\sin \delta$. No current is observed if the system is
symmetric $\delta = n \pi$. This is the underdamped analog of a
similar effect which occurs for {\em overdamped} motion in systems
with state dependent diffusion \cite{Buttiker,Kampen,Landauer1}. In
the overdamped case, directed motion appears because a particle
diffusing up a potential hill can easier surmount a potential barrier
if the ascent occurs in a heated region \cite{Landauer2}. In contrast,
in the presence of a very small frictional force, a particle
experiences the microscopic temperature modulation many times, both
when the particle is in a well and even after its escape into a
running state. Thus, the current must vanish in the zero dissipation
limit.    

The motion of a classical particle in an external potential
and subject to thermal noise is governed by Kramers'
equation\cite{Kramerspr}. While the overdamped motion (Smoluchowski
limit) is relatively easy to handle, the underdamped limit is much
more complex. The steady state  
\begin{figure}
\narrowtext
{\epsfxsize=5.0cm\epsfysize=5.0cm\centerline{\epsfbox{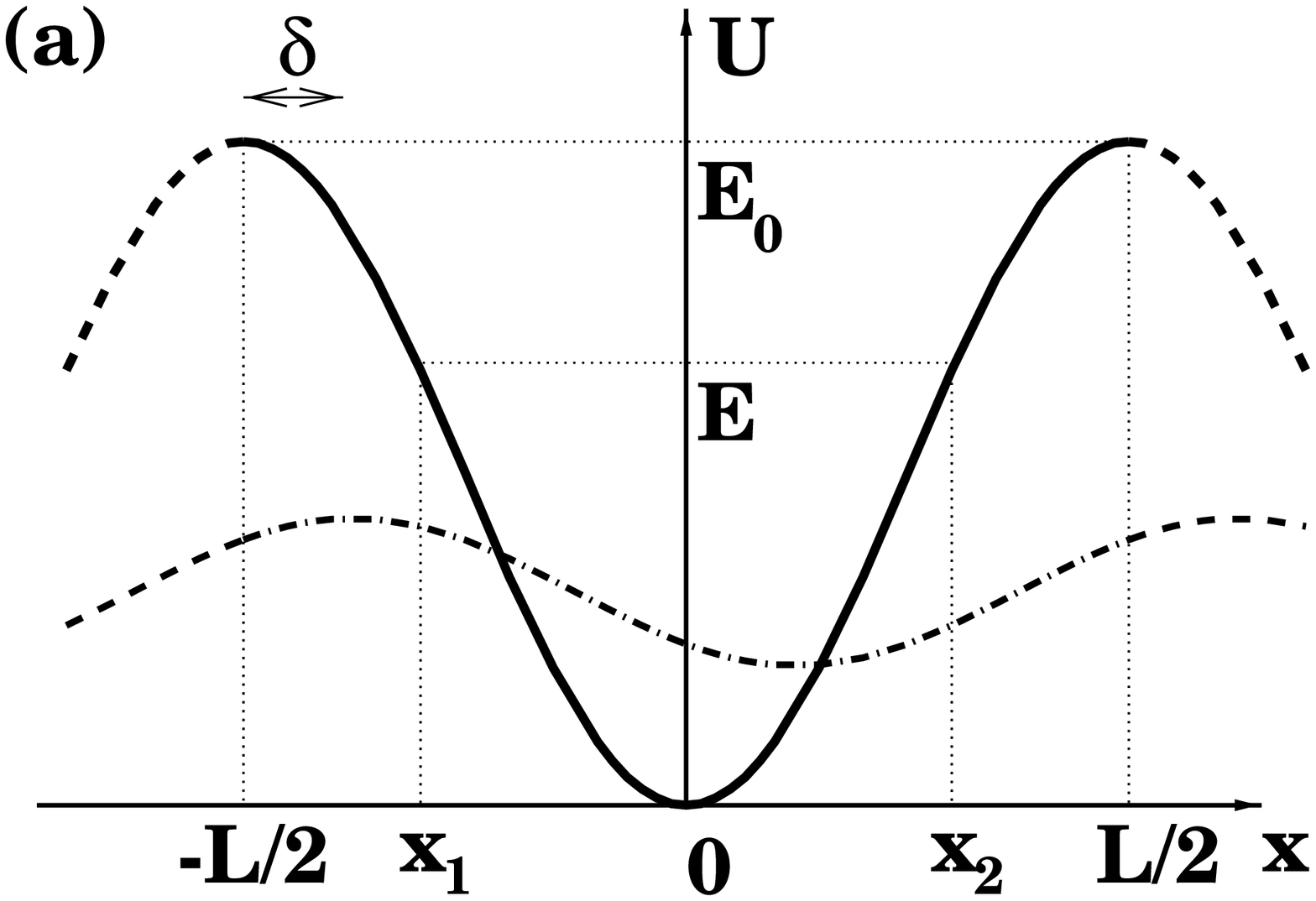}}}
{\epsfxsize=5.0cm\epsfysize=5.0cm\centerline{\epsfbox{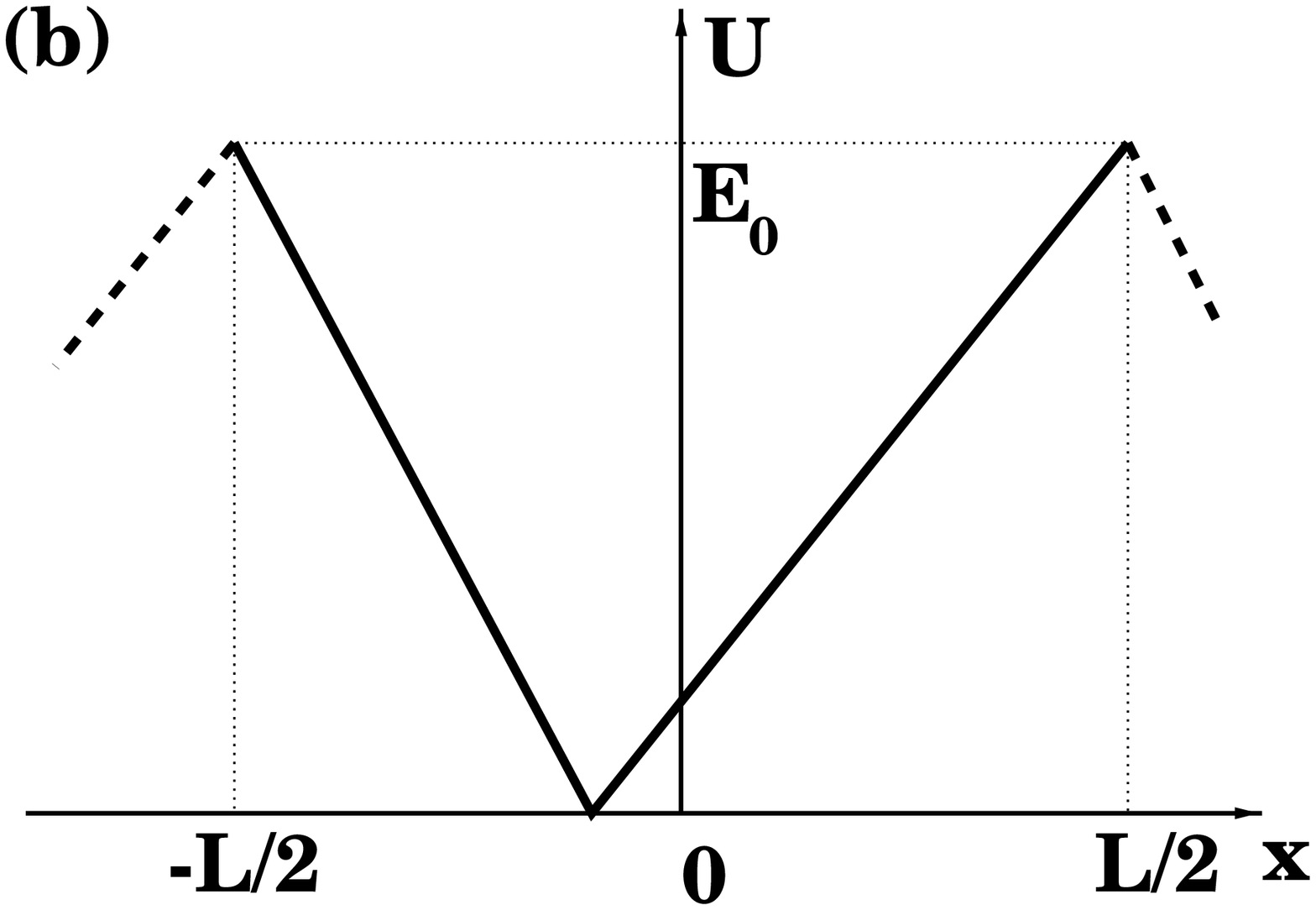}}}

\protect\vspace{0.5cm}

\caption{One period of the potential $U(x)$, (a) the
potential given by Eq.(\protect\ref{pot}) (bold solid line),
temperature variation $T(x)$ (dash-dotted line);  
(b) an example of sawtooth potential.} 
\label{fig1}
\end{figure}
\noindent
current of particles 
in a periodic potential due to a constant external force was
treated by Risken and Vollmer \cite{RV}. Here we generalize their
approach to the case of a spatially non-uniform temperature, thus
presenting a rare example of a solution of an underdamped
problem. Kramers' equation \cite{Kramerspr} for the joint distribution
function $W(x,v)$ of coordinate and velocity for a
particle with mass $m$ in the potential
$U(x)$ subject to a frictional force with damping constant $\gamma$
and a temperature $T(x)$ is 
\begin{equation}\label{FP}
-mv\frac{\partial W}{\partial x} + \frac{\partial U}{\partial
x}\frac{\partial W}{\partial v} + \gamma \frac{\partial}{\partial v}
\left[ \frac{T(x)}{m} \frac{\partial W}{\partial v} + vW \right] = 0. 
\end{equation} 
We assume that both $U(x)$ and $T(x)$ are periodic
functions with the same period $L$, and the potential $U(x)$ has only
one maximum per oscillation period. The amplitude of
the potential is $E_0$, the zero of energy is chosen at the minima of
the potential, and the points $x = \pm L/2$ correspond to the
potential maxima (Fig.~1). Furthermore we assume that $T(x) \ll E_0$
for any $x$.  

Eq. (\ref{FP}) must be supplemented by boundary conditions. For $E >
E_0$ the distribution function is periodic in $x$, $W(-L/2,v) =
W(L/2,v)$. For energies below $E_0$ all particles are reflected from
the potential barrier at the points $x_1(E)$ and $x_2(E)$
(Fig.~1). Thus, for $E < E_0$, the boundary condition requires that at
the turning points the distribution function of incoming particles
is the same as that of the reflected particles. 

Following Ref. \cite{RV}, we replace the velocity $v$ by the energy
variable $E = mv^2/2 + U(x)$. It is then necessary to discriminate
between left- and right-moving particles; their distribution functions
are denoted as $W^-$ and $W^+$, respectively. Introducing for further
convenience the symmetric and antisymmetric combinations, $W^{S,A} =
W^+ \pm W^-$, we obtain \cite{RV}
\begin{eqnarray} 
\frac{\partial W^A}{\partial x} & = & \gamma \frac{\partial}{\partial
E} \left\{ v(x,E) \left[ 1 + T(x) \frac{\partial}{\partial E} \right]
W^S (x,E)\right\}; \label{FP1} \\ 
\frac{\partial W^S}{\partial x} & = & \gamma \frac{\partial}{\partial
E} \left\{ v(x,E) \left[ 1 + T(x) \frac{\partial}{\partial E} \right]
W^A (x,E)\right\}, \label{FP2}
\end{eqnarray} 
with the boundary conditions
\begin{eqnarray} \label{bound1}
\matrix{
W^{S,A} (-L/2,E) = W^{S,A} (L/2,E)\ , & E > E_0 \cr
W^{A} (x_1(E),E) = W^{A} (x_2(E),E) = 0\ , & E < E_0 } .
\end{eqnarray}
Here $v(x,E) = [(2/m)(E - U(x))]^{1/2}$, and the turning points
$x_{1,2} (E)$ are the solutions of the equation $U(x) = E$. In
addition, both functions $W^S$ and $W^A$ and their energy derivatives
are continuous for $E = E_0$. The function $W^S$ is normalized
according to 
\begin{equation} \label{norm}
\int dE dx \left[mv(x,E)\right]^{-1} W^S (x,E) = 1, 
\end{equation} 
and the expression for the (particle) current has the form 
\begin{equation} \label{curdef}
I = m^{-1} \int dE W^A (x,E).
\end{equation}
Integrating Eq. (\ref{FP1}) over energy, we conclude immediately that
the current $I$ does not depend on $x$. Note that no assumption about
damping has been made so far --- equations (\ref{FP1}),
(\ref{FP2}) are valid for arbitrary $\gamma$.   

Now we turn to the underdamped case. For $\gamma = 0$, both
the symmetric and the antisymmetric distribution functions $W^S$ and
$W^A$ depend on energy only. Then due to the boundary
conditions $W^A$ vanishes for $E < E_0$, and the matching conditions
for $E = E_0$ imply that it vanishes also for $E > E_0$. Thus in the
absence of friction there is no current \cite{foot1}. 
Dissipation is essential for the 
migration of particles between different potential wells.
For low dissipation the typical time for this migration
is much longer than the period of the oscillations inside each well,
and thus the energy varies much slower in time than the coordinate $x$
of the particle. Consequently, the distribution function depends weakly
on $x$ while it varies rapidly with energy. This rapid dependence is, 
however, essential only in the narrow layer of energies around $E =
E_0$; the thickness of this layer is proportional to a positive power
of $\gamma$. Generalizing the approach of Ref. \cite{RV} we write
\begin{equation} \label{Anzatz}
W^{S,A} (x,E) = \tilde W^{S,A} (x,E) + w^{S,A}(x,E),
\end{equation}
where $\tilde W^{S,A}$ are slow functions of $x$ and $E$, while
$w^{S,A}(x,E)$ are slow functions of $x$ that vary rapidly with
energy. The functions $w^{S,A}$ are different from zero only in the
narrow layer of energies around $E_0$; their role is to ensure the
continuity of $W^{S,A}$ for $E = E_0$.
For $E > E_0$, the boundary conditions read now
\begin{eqnarray} \label{bound2}
\tilde W^{S,A} (-L/2,E) = \tilde W^{S,A} (L/2,E); \nonumber \\
w^{S,A} (-L/2, E) = w^{S,A} (L/2,E),
\end{eqnarray}
and for $E < E_0$
\begin{equation} \label{bound3}
\tilde W^A (x_{1,2}(E),E) = w^A (x_{1,2}(E),E) = 0.
\end{equation}

First we consider the functions $\tilde W^{S,A}
(x,E)$. The periodic functions $v(x,E)$, $t(x,E) \equiv v(x,E)T(x)$
and the distributions $\tilde W^{S,A}(x,E)$
can be expanded in Fourier series (in the range $E < U(x)$ we define
them to be zero),   
\begin{eqnarray*}  
v(x,E) & = & v_0 + \sum_{n=1}^{\infty} v_{nc} \cos \frac{2\pi nx}{L} +
\sum_{n=1}^{\infty} v_{ns} \sin \frac{2\pi nx}{L}; \\
t(x,E) & = & t_0 + \sum_{n=1}^{\infty} t_{nc} \cos \frac{2\pi
nx}{L} + \sum_{n=1}^{\infty} t_{ns} \sin \frac{2\pi nx}{L}; 
\end{eqnarray*}
\begin{eqnarray*} 
\tilde W^{S,A}(x,E) & = & \tilde W^{S,A}_0 (E) + \sum_{n=1}^{\infty}
\tilde W^{S,A}_{nc} (E) \cos \frac{2\pi nx}{L} \\
& + & \sum_{n=1}^{\infty} \tilde W^{S,A}_{ns} (E) \sin \frac{2\pi nx}{L}.
\end{eqnarray*}
It is seen from Eqs. (\ref{FP1}), (\ref{FP2}) that the coefficients
$\tilde W^{A}_{nc,s}$ are proportional to $\gamma$. Then for $E < E_0$
due to the boundary condition the whole function $\tilde W^A$ is
proportional to $\gamma$, and consequently the $x$-dependent part of
$\tilde W^S$ is proportional to $\gamma^2$. In the following we
restrict our consideration to the terms proportional to $\gamma$, thus
neglecting the coefficients $\tilde W^{S}_{nc,s}$. Taking the
constant part of Eq. (\ref{FP1}), we find for $E < E_0$
\begin{equation} \label{func1}
\tilde W^S (E) \equiv N P(E) =  N \exp \left\{ - \int^E_{E_0} \frac{v_0
(E')}{t_0 (E')} dE' \right\},
\end{equation}
with a normalizing constant $N$. With the 
abbreviation $\phi (x,E) = 1 - ({v_0 (E)}/{t_0 (E)})
T(x)$ the solution for $\tilde W^A$ satisfying the boundary condition
for $E < E_0$ is   
$$\tilde W^A (x,E) = \gamma \frac{\partial}{\partial E} \left\{ \tilde
W^S (E) \int_{x_1(E)}^{x} v(x',E)  \phi (x',E) dx' \right\}. 
$$
We note that for potentials $U(x)$ (such as that of Fig.~1a) with a
saddle at $E_0$, this expression diverges as $E \to E_0$ and $x
\to \pm L/2$. This divergence is similar to that of the oscillation
period for $E \to E_0$. It is not detrimental, since the Jacobian from
$(x,v)$ to the $(x,E)$ space has a compensating singularity. It is
advantageous to treat this divergence by considering a sawtooth potential, 
like that of Fig.~1b, for which all expression converge. We will
proceed in this way, and our final results will be shown to be
convergent in general case. Thus, we expect that our
consideration is also valid for potentials with a saddle.  

For $E > E_0$ the function $\tilde W^S$ still does not depend on $x$
(up to  terms of order $\gamma$), and is given by Eq. (\ref{func1}). For
$\tilde W^A (x,E)$ we find
\begin{eqnarray} 
\tilde W^A (x,E) & = & \gamma \frac{\partial}{\partial E} \left\{ \tilde
W^S (E) \int_{-L/2}^{x} v(x',E) \phi (x',E) dx' \right\} \nonumber \\
& + & B(E), 
\label{func3}
\end{eqnarray}
\noindent
which differs from the solution for $E < E_0$ by the
position-independent function $B(E)$. The latter appears due to the
different boundary conditions, and is determined by
Eq. (\ref{FP2}). Taking the constant part of this equation and fixing
the only integration constant so that $B(E_0) = 0$, we obtain
\begin{eqnarray} \label{func4}
B(E) & = & P(E) \left[ \int_{E_0}^E \frac{F(E')}{P(E')} dE' + G(E_0)
\right] - G(E). 
\end{eqnarray}
Here we have introduced the functions
\begin{eqnarray*}
F(E) & = & -\frac{1}{2t_0(E)} \sum_{n=1}^{\infty} \sum_{i = c,s}
\left[ v_{ni} (E) \tilde W^A_{ni} + t_{ni}(E)\frac{\partial \tilde
W^A_{ni}}{\partial E} \right],  \\
G(E) & = & -\frac{\gamma}{L} \frac{\partial}{\partial E} \left\{ \tilde
W^S (E) \int_{-L/2}^{L/2} x v(x,E) \phi (x,E) dx \right\} .
\end{eqnarray*}
The functions $\tilde W^A_{nc,s}$ (which are just the coefficients 
in the Fourier series for the solution Eq. (\ref{func3})) are easily 
obtained from Eq. (\ref{FP1}) and read 
\begin{eqnarray*}
\tilde W^A_{nc,s} = \pm \frac{\gamma L}{2\pi n}
\frac{\partial}{\partial E} \left\{ \left[ v_{ns,c} (E) + t_{ns,c} (E)
\frac{\partial}{\partial E} \right] \tilde W^S (E) \right\}.
\end{eqnarray*}
With the help of the antisymmetric product, 
$$\left\{ a \times b \right\} = (2L^2)^{-1} \int_{-L/2}^{L/2} a(x)
b(x') sign (x-x') dx dx',$$
the expression for $B(E)$ after some algebra can be brought into the
form   
\end{multicols}
\widetext
\vspace*{-0.2truein} \noindent \hrulefill \hspace*{3.6truein}
\begin{eqnarray} \label{func5}
B(E) = -\gamma L N P(E) \int_{E_0}^E \frac{dE'}{t_0 (E')} \left[
\left\{ v \times \frac{\partial v}{\partial E'} \right\} -
\frac{v_0}{t_0} \left\{ v \times \frac{\partial t}{\partial E'}
\right\} - 2\frac{v_0}{t_0} \left\{ t \times \frac{\partial
v}{\partial E'} \right\} + 2\frac{v_0^2}{t_0^2} \left\{ t \times
\frac{\partial t}{\partial E'} \right\} \right].
\end{eqnarray}
\hspace*{3.6truein}\noindent \hrulefill 
\begin{multicols}{2}
\noindent
Now we cure the discontinuity of $\partial \tilde W^A/\partial E$ for
$E = E_0$ by taking into account the functions $w^{S,A} (x,E)$. 
Since they vary rapidly with energy, the
derivatives $\partial w/\partial E$ can be eliminated in favor of
$\partial^2 w/\partial E^2$. Replacing $E$ by $E_0$ in the functions
which are smooth in the energy, neglecting the difference between
$x_1(E)$ and $-L/2$, and introducing the dimensionless variable $u$, 
$$u(x) = -\pi + 2\pi \frac{\int_{-L/2}^x T(x) v(x,E_0)
dx}{\int_{-L/2}^{L/2} T(x) v(x,E_0) dx}$$ 
(which is an analog of the action for the uniform temperature case), $-\pi
\le u \le \pi$, we find   
\begin{equation} \label{contin}
\frac{\partial w^{S,A}}{\partial u} = u_0 \frac{\partial^2
w^{A,S}}{\partial E^2}.
\end{equation}
Here $u_0 = \gamma L t_0 (E_0)/(2\pi)$, and 
the boundary conditions read $w^A (\pm \pi, E) = 0$ for $E < E_0$, and
$w^{S,A} (-\pi, E) = w^{S,A} (\pi, E)$ for $E > E_0$. 
 
Eq. (\ref{contin}) has been investigated by
Risken and Vollmer \cite{RV}. Making use of their solution, writing
down the matching conditions for $W^A$ at $E = E_0$, and taking into
account that the function $B(E)$ in Eq. (\ref{func3}) is proportional
to $\gamma$, we find that the functions $w^{S,A}$ are proportional to
$\gamma^{3/2}$. Thus, these functions are not essential for our
consideration. For the current we obtain then 
\begin{equation} \label{cur2}
I = m^{-1} \int_{E_0}^{\infty} B(E) dE.
\end{equation}   
Eqs. (\ref{func5}) and (\ref{cur2}) together with the definition
(\ref{norm}) of the normalizing constant $N$ yield a closed
expression for the current induced by an arbitrary temperature
distribution $T(x)$. This current is proportional to $\gamma$ and
vanishes in the limit of constant temperature $T(x) = T_0$, since then
$t(x,E) = T_0 v(x,E)$. 

We continue with two simplifying assumptions. First, we consider
the experimentally relevant regime of small temperature
gradients. Specifically, following Ref. \cite{Buttiker}, we assume 
$$T(x) = T_0 + T_1 \cos (2\pi x/L - \delta), \ \ \ \ T_1 \ll T_0.$$
Furthermore, for simplicity, we choose the potential $U(x)$ to be
symmetric, $U(x) = U(-x)$. To linear order in $T_1$,
the expression for the current simplifies and is given by 
\begin{eqnarray} \label{cur3}
I & = & \frac{\gamma L N T_1}{m T_0 L^2 v_0 (E_0)} \sin \delta
\int_{-L/2}^{L/2} dx \sin \left( \frac{2\pi x}{L} \right)
\frac{\partial v(x,E_0)}{\partial E} \nonumber \\
& \times & \int_0^x v(x', E_0) dx',
\end{eqnarray}
where we used the condition $T_0 \ll E_0$. As is expected, the current
is proportional to $\sin \delta$, and vanishes when both the potential
and the temperature are symmetric. We also note that Eq. (\ref{cur3})
is finite for a potential $U(x)$ of a general form (even though
$\partial v(x, E_0)/\partial x$ diverges for $x \to \pm L/2$), and
thus, we expect that it is generally valid. As an example, we consider
the simple form of $U(x)$,  
\begin{equation} \label{pot}
U(x) = ({E_0}/{2}) \left [1 - \cos ({2\pi x}/{L}) \right], 
\end{equation}
shown in Fig.~1a. We obtain
\begin{equation} \label{cur4}
I = \frac{\gamma T_1}{2m T_0} \exp \left( -\frac{E_0}{T_0} \right)
\sin \delta.
\end{equation}
It is instructive to compare this expression with the one found in the
overdamped regime \cite{Buttiker,Kampen} (index $ov$). 
For the same parameters we find
\begin{equation} \label{cur5}
I_{ov} = \frac{\pi^2 E_0^2 T_1}{\gamma T_0^2 L^2} \exp \left(
-\frac{E_0}{T_0} \right) \sin \delta.
\end{equation}
First, we note that the sign of the current is the same in the overdamped
and in the underdamped regime. It corresponds to the notion
\cite{Landauer2,Buttiker} that it is easier for the particle to climb a
``hot'' slope. 
Furthermore, we note that the two expressions (\ref{cur4}) and
(\ref{cur5}) match for a certain critical damping,
\begin{equation} \label{dampcr}
\gamma_c \sim ({E_0}/{L}) ({m}/{T_0})^{1/2}.
\end{equation} 
This critical damping can be understood by considering the overdamped 
side. Indeed, to leading order, the overdamped approximation
\cite{Buttiker,Kampen} neglects the terms which do not contain
$\gamma$ in Eq. (\ref{FP}). This is justified for $\gamma \gg v^{-2}
\partial U/\partial x$, where $v \sim (T/m)^{1/2}$ is the typical
thermal  velocity. Substituting $E_0/L$ for the characteristic value
of the potential derivative, we readily obtain
Eq. (\ref{dampcr}). Thus, we conclude that our expression for the
underdamped regime (\ref{cur4}) is valid up to $\gamma \sim \gamma_c$.

Now we compare the result (\ref{cur4}) with the current $I \sim
(F/\gamma L) \exp(-E_0/T_0)$, obtained in Ref. \cite{RV} for the
underdamped Brownian motion due to an external force $F$. We see that
a non-uniform temperature has the same effect as an applied effective
uniform driving force $F_{eff} \sim \gamma^2 L T_1 \sin
\delta/(mT_0)$. Thus, the motion generated by a non-uniform
temperature profile is less effective than the forced motion due
to a uniform field. This is in contrast to the overdamped case
\cite{Buttiker}, where the effective force $F_{eff} \sim E_0
T_1 \sin \delta /(T_0 L)$ does not depend on $\gamma$.  

In conclusion, we investigated how a non-uniform temperature can produce
a steady, directed current of particles in a periodic potential 
in the underdamped regime. A current is generated even in the case of 
a symmetric potential provided the temperature profile is asymmetric 
(out of phase). A temperature oscillation is equivalent to an external 
driving force with a magnitude proportional to the square of the friction
constant.

The work was supported by the Swiss National Science Foundation.

\end{multicols}

\begin{references}

\bibitem{JAP} F.~J\"ulicher, A.~Ajdari and J.~Prost,
Rev. Mod. Phys. {\bf 69}, 1269 (1997); R. D. Astumian, Science, {\bf
276}, 917 (1997).  

\bibitem{motor} R.~D.~Vale and F.~Oosawa, Adv. Biophys., Ed. by
M.~Kotani (Elsevier, Limerick, 1990) {\bf 26}, 97;  A.~Ajdari and
J.~Prost, C. R. Acad. Sci. Paris II {\bf 315}, 1635 (1992).  
      
\bibitem{Feynm} R. P. Feynman, R. B. Leighton and M. Sands, 
{\it The Feynman Lectures in Physics} (Addison-Wesley, Reading,
1966), Vol. 1, Chap. 46.

\bibitem{Buttiker} M.~B\"uttiker, Z. Phys. B {\bf 68}, 161 (1987).

\bibitem{Kampen} N.~G.~van Kampen, IBM J. Res. Dev. {\bf 32}, 107
(1988). 

\bibitem{Landauer1} R.~Landauer, J. Stat. Phys. {\bf 53}, 233 (1988). 

\bibitem{Millonas} A model equivalent to the application of spatial
dependent temperature was derived from the microscopic classical
description of a system coupled  to a bath of oscillators, see
M.~M.~Millonas, Phys. Rev. Lett. {\bf 74}, 10 (1995); {\em ibid}, {\bf
75}, 3027E (1995); A.~M.~Jayannavar, Phys. Rev. E {\bf 53}, 2957
(1996).  

\bibitem{force} M.~O.~Magnasco, Phys. Rev. Lett. {\bf 71}, 1477
(1993); R.~Bartussek, P.~H\"anggi, and J.~G.~Kissner,
Europhys. Lett. {\bf 28}, 459 (1994); M.~M.~Millonas and M.~I.~Dykman,
Phys. Lett. A {\bf 185}, 65 (1994); C.~R.~Doering, W.~Horsthemke, and
J.~Riordan, Phys. Rev. Lett. {\bf 72}, 2984 (1994); I.~Der\'enyi and
T.~Vicsek, {\em ibid}, {\bf 75}, 374 (1995). 

\bibitem{fluct} R.~D.~Astumian and M.~Bier, Phys. Rev. Lett. {\bf 72},
1766 (1994).  

\bibitem{Prost} J.~Prost, J.-F.~Chauwin, L.~Peliti, and A.~Ajdari,
Phys. Rev. Lett. {\bf 72}, 2652 (1994).

\bibitem{inertia} P.~Jung, J.~Kissner, and P.~H\"anggi, 
Phys. Rev. Lett. {\bf 76}, 3436 (1996). 

\bibitem{quant1} S.~Yukawa, M.~Kikuchi, G.~Tatara, and H.~Matsukawa,
J. Phys. Soc. Jap. {\bf 66}, 2953 (1997);
P.~Reinmann, M.~Grifoni, and P.~H\"anggi, Phys. Rev. Lett. {\bf 79},
10 (1997); 
G.~Tatara, M.~Kikuchi, S.~Yukawa, and H.~Matsukawa,
cond-mat/9711045 (unpublished); I.~A.~Goychuk, E.~G.~Petrov, and
V.~May, Phys. Lett A {\bf 238}, 59 (1998). 

\bibitem{Landauer2} Bistable systems with state dependent diffusion
have long been of interest to demonstrate that in non-equilibrium
systems the relative stability of locally stable states is affected by
the kinetics even far from these states. See R. Landauer, Phys. Rev. A
{\bf 12}, 636 (1975).  

\bibitem{Kramerspr}  H. A. Kramers, Physica {\bf 7}, 284 (1940). 

\bibitem{RV} H.~Risken and H.~D.~Vollmer, Z. Phys. B {\bf 35}, 177
(1979). 

\bibitem{foot1} This is in contrast to the system driven by a constant
force, see {\em e.g.} \protect\cite{RV}. This force breaks the
symmetry between $v$ and $-v$, implying an infinite current for vanishing
dissipation. 

\end{references}
\end{document}